\documentclass{modagujournal}
\draftfalse

\journalname{JGR-Oceans}

\begin{document}

%
%


\title{Characterization of the structure and cross-shore transport properties of a
coastal upwelling filament using three-dimensional finite--size Lyapunov exponents}

%
%




\authors{Jo\~ao H. Bettencourt\affil{1}\thanks{Current address, LEGOS-OMP, 18 Av. Edouard Belin, Toulouse, France},
Vincent Rossi\affil{1}, \\
Emilio Hern\'andez-Garc\'ia\affil{1},\\
Martinho Marta-Almeida\affil{2},\\
and Crist\'obal L\'opez\affil{1}}


\affiliation{1}{IFISC(CSIC-UIB), \\
	Instituto de Fisica Interdisciplinar y Sistemas Complejos.\\
	Campus Universitat de les Illes Balears. \\
	E-07122 Palma de Mallorca, Spain}

\affiliation{2}{Instituto Espa\~{n}ol de Oceanograf\`{i}a, \\
	Centro Oceanogr\'{a}fico A Coru\~{n}a,\\
	15001 A Coru\~{n}a, Spain}




\correspondingauthor{Jo\~ao H. Bettencourt}{jhbettencourt@protonmail.com}




\begin{keypoints}
\item Three-dimensional structure \added{and hydrographic characteristics} of \added{an} upwelling filament \added{are} analysed using Lagrangian tools.
\item Low values of finite--size Lyapunov exponent evidence low dispersion \added{and relatively homogeneous thermohaline properties}
inside \added{the} filament.
\item Lagrangian Coherent Structures delimit the filament, isolate it from surrounding waters \added{and match hydrographic gradients.}
\end{keypoints}

%
%


\begin{abstract}
The three dimensional structure, dynamics and dispersion characteristics
of a\deleted{n} \added{simulated} upwelling filament
in the Iberian upwelling system are analyzed using Lagrangian tools. \added{We used a realistic} \deleted{applied to a} regional simulation of the western Iberian shelf \deleted{using realistic forcings,} which is \replaced{coincident}{concomitant}  with an \emph{in-situ} \added{oceanographic} campaign that \replaced{studied}{surveyed} the \deleted{oceanography of the} area.
\replaced{We compute 3d
fields of finite--size Lyapunov exponents (FSLE) from 3d velocity fields and extract the field's
ridges to obtain proxies to the Lagrangian structures that form the boundaries
of a cold water
filament that develops due to the interaction of a mesoscale eddy with the upwelling front.
The cold, upwelled waters move along the filament, conserving their density. The filament
itself is characterized by small dispersion of fluid elements in its interior.
The comparison with potential temperature gradient  fields shows that the limits of
the filament coincide with large gradient regions, which explain the
isolation of the interior of the filament from the external waters.We conclude that the
Lagrangian analysis used in this work is useful in explaining the dynamics of across shore exchanges of material between coastal regions
and the open ocean due to mesoscale processes.}{We compute 3d fields of finite--size Lyapunov exponents (FSLE) from 3d velocity fields and extract the field's ridges to study the spatial distribution and temporal evolution of the Lagrangian Coherent Structures (LCSs) evolving around the filament. We find that the most intense curtain-like LCSs delimit the boundaries of the whole filamentary structure whose general properties match well the observations. The filament interior is characterized by small dispersion of fluid elements. Furthermore, we identify a weak LCS separating the filament into a warmer vein and a colder filament associated with the interaction of a mesoscale eddy with the upwelling front. The cold upwelled water parcels move along the filament conserving their density. The filament itself is characterized by small dispersion of fluid elements in its interior. The comparison of LCSs with potential temperature and salinity gradient fields shows that the outer limits of the filament coincide with regions of large hydrographic gradients, similar to those observed, explaining the isolation of the interior of the filament with the surrounding waters. We conclude that the Lagrangian analysis used in this work is useful in explaining the dynamics of cross-shore exchanges of materials between coastal regions and the open ocean due to mesoscale processes.}

\end{abstract}

%
%

%


%
%
%
%

\section{Introduction}
\label{sec:c6-intro}

Mesoscale filamental structures are ubiquitous features of the ocean and
more particularly in the coastal ocean, in which they play a key role in cross-shelf exchanges
of water masses. In upwelling regions this role
of filaments is of even a greater importance due to the intense biological productivity near-shore and their fertilizing role of the adjacent oligotrophic ocean \citep{Alvarez2007}.
In the Iberian Peninsula Upwelling System, mesoscale processes, rather than large-scale
variability, are the dominant factors controlling the strength of the coastal
upwelling \citep{Relvas2009} and the response of
its associated ecosystem \citep{Rossi2013}.
The regional oceanography reveals  a series of mesoscale
structures such as jets, meanders, eddies and upwelling filaments,
superimposed on the large--scale seasonal patterns
\citep{Peliz2002}. From a biological perspective,
dynamics with spatial scales of 10-100 km are determinant for the control of larval
retention and dispersal \citep{queiroga2007} and more generally, of the distribution and variability of marine planktonic communities \citep{Relvas2007,Cravo2010,Rossi2013,hernandez2014reduction}.

In the late Spring/early Summer, predominantly northerly coastal winds start to blow in
the western Iberian Peninsula, driving an offshore Ekman transport in the surface layers that
forces the upwelling of colder, nutrient rich, subsurface waters along the coast. The depth of the
surface Ekman layer usually ranges 20-60 m and the upwelled waters come from as deep as 200 m in a 10-20 km
wide strip of coastal upwelling with vertical velocities of the order of $\sim$ 10 m day\textsuperscript{-1}
\citep{Barton2001b}. Shortly after the initiation of the upwelling--favorable winds, large filaments
start to develop, associated with strong offshore currents of $\sim$ 0.5 $m/s$
that can extend more than 200 km \citep{Haynes1993,Relvas2007,Rossi2013}. The significant offshore
mass transport along the major axis of a filament is larger than that possible by purely wind-driven Ekman
dynamics \citep{Relvas2007}.
Calculations by \citep{Rossi2013} have suggested that filaments are responsible for greater offshore transport of coastal properties in the Iberian system ($\sim 60$\% of total cross-shelf exchanges) and carbon export by filaments is 2.5 to 4.5 times that due to Ekman transport \citep{Alvarez2007}. Overall,
filaments provide an important mechanism for exchange of materials between coastal and
offshore waters, especially the transport of chlorophyll \citep{Rossi2013}, nutrients \citep{Cravo2010}
and carbon \citep{SantanaFalcon2016}.


\emph{In-situ} observations in Western Iberia have elucidated some aspects of the physical and biological
oceanography of upwelling filaments. The offshore flow was found to be limited to a relatively thin surface
layer, with the strongest flow observed at the boundaries of the structure, while substantial
reverse onshore flow may occur beneath the filament \citep{Barton2001a,Rossi2013}. Weak mixing in the core of
filaments but enhanced  at their boundaries has been reported \citep{Barton2001a}. Note that close to barriers there are strong gradients of transported substances, so that mixing is enhanced in their vicinity. Water within filaments
has been found to be relatively homogeneous but well isolated from the surroundings \citep{Rossi2013}.
Mesoscale eddies near the upwelling front,  resulting from the interactions of the poleward subsurface flow with the surface upwelling
jet and the topography, seem to contribute to the filament development \citep{Peliz2002}. Indeed, on some
occasions, opposite rotating eddies are observed at the base and the tip of the filament, creating typical mushroom-like
structures and promoting its generation and fast offshore development. Lastly, the riverine coastal low salinity plume often
observed off north-west Iberia \citep{Otero2010} has been mentioned as a possible input of buoyancy to the filamentary
structure, possibly promoting its offshore elongation \citep{Peliz2002,Rossi2013}. A complete description of
\replaced{filaments}{the filaments'}
characteristics and a better understanding of their dynamics is required to properly assess their role in cross-shelf
exchanges.

In this paper we use Lagrangian tools to study the three-dimensional structure of a coastal cold water
filament in a regional simulation of the western Iberian shelf (and adjacent ocean) using realistic forcings.
This simulation is coincident with an \emph{in-situ} campaign that studied the bio-physical
oceanography of the region under upwelling-favorable conditions \citep{Rossi2010,Rossi2013}. Hence, our numerical
\replaced{analyses}{analysis} deepens, complements and generalizes that observational study, while, conversely, some aspects of the
numerical results are validated by the \emph{in-situ} results of \citet{Rossi2013}.
In particular they documented peaks of biological activity associated with the successive upwelling pulses and reported that the offshore transport of enriched, recently-upwelled waters, is more efficient
through the filament than across the front. The precise reasons behind such an efficient transport within this
apparently coherent filamentary structure remained puzzling and motivated the present paper.

We compute 3d fields of finite-size
Lyapunov exponents (FSLE) and extract the 3d boundaries of the filament as ridges of the FSLE fields. Virtual particles
are released and tracked numerically in order to characterize the dynamics of the water masses inside the filament\deleted{ and to compute offshore heat
and salinity fluxes due to the filament}.
Section \ref{sec:c6-data} describes the data and methods used in this work, while Section \ref{sec:c6-results} presents
and discusses the results. Section \ref{sec:discussion} discusses
the results and conclusions are drawn in Section \ref{sec:c6-conclusions}.

\section{Data and methods}\label{sec:c6-data}

\subsection{Area and period of study}

To allow validating and complementing our numerical approach with observations, we focus on
the central region of the Iberian upwelling in late summer 2007. This is concomitant with the MOUTON07 survey
which studied the physical and biogeochemical
variability of the coastal upwelling and of its connection with the open ocean.

More details about the MOUTON07 campaign and dataset are reported in the literature. In particular,
\citet{Rossi2010} observed a secondary "shelf-break" upwelling front, in addition to the common coastal one,
and proposed a physical mechanism for its generation with the use of numerical modelling. \citet{Rossi2013}
investigated the cross-shore variability and transport in the central Iberian upwelling by comparing two zonal
sections across the front and an extensive sampling of an upwelling filament originating at about 40.3$^o$N and
developing through September.

We thus focus our numerical analysis on the central Iberian upwelling in September 2007. This period still falls within the upwelling season but it experienced low to moderate northerly winds, hence favouring the formation and maintenance of
well-defined filaments.

\begin{figure}
	\centering
    \includegraphics[width=0.7\textwidth]{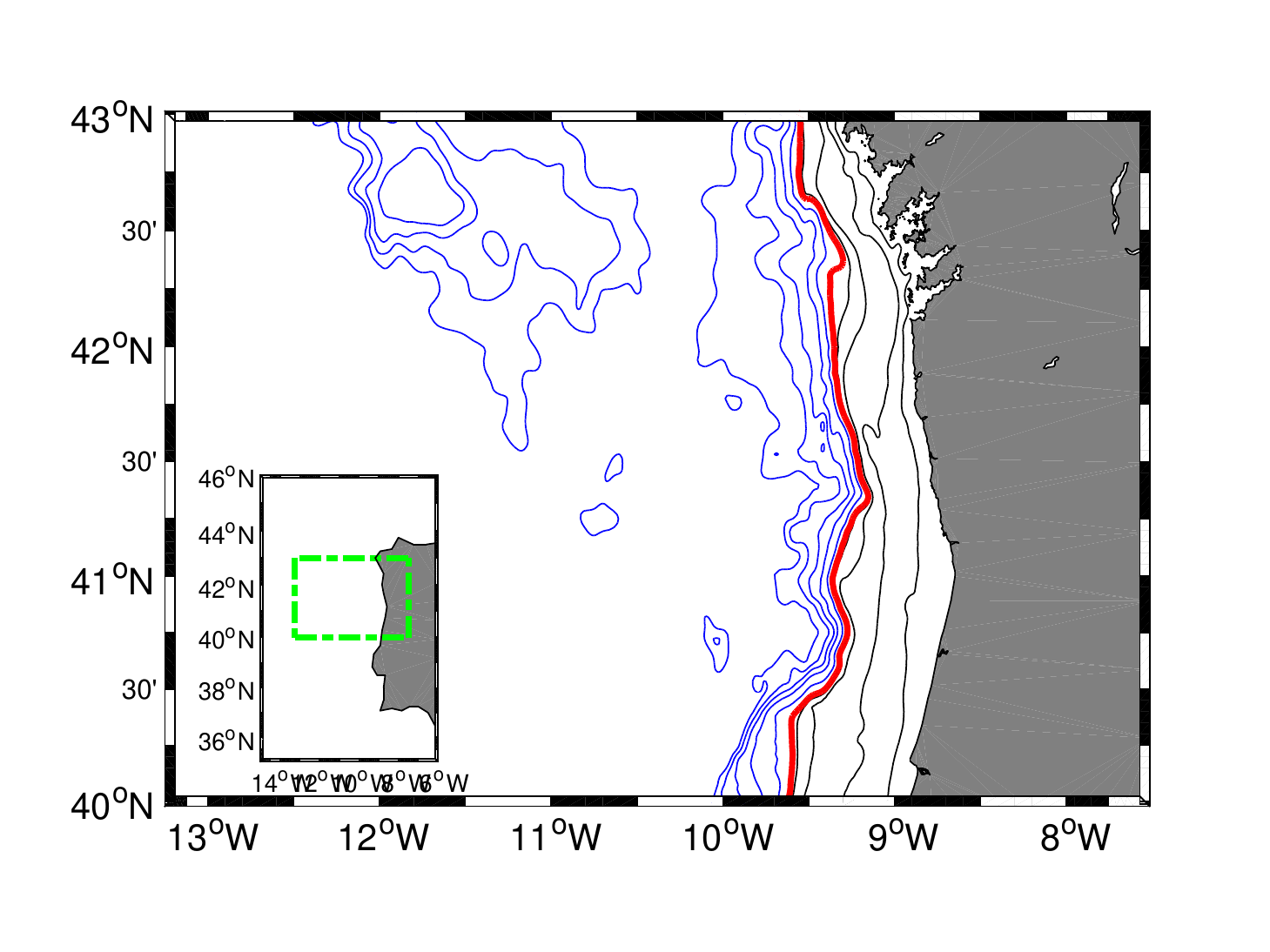}
	\caption{Region of calculation of the 3d FSLE fields. Thick red line: 250 m isobath; Black lines: 50, 100, 150, 200 and 250 m isobaths; Blue lines: isobaths between 500 and 2500 m, drawn every 500 m. Inset:
		ROMS hindcast region and FSLE region (green rectangle).}
	\label{fig:west-iberia}
\end{figure}

\subsection{Simulations of the Iberian upwelling}


The oceanic simulations were done using the Regional Ocean
Modeling System (ROMS, \citep{shchepetkin_2005}).
ROMS is a free-surface terrain-following model which solves the
primitive equations using the Boussinesq and hydrostatic approximations.
This state-of-the-art model is highly configurable for realistic
applications and has been applied to a wide variety of space and time
scales across the world \citep{haidvogel_2008}.

The model domain covers the Portuguese and Galician region,
extending well beyond the shelf break (13.25$^o$W to 7.5$^o$W
and 35.6$^o$N to 45$^o$N) with horizontal resolution of 2 km in the meridional
direction and 1 km in the zonal direction (see Fig. \ref{fig:west-iberia}). The vertical discretization
used 30 $\sigma$-layers, stretched to increase resolution near the
surface and bottom. The bathymetry is interpolated from ETOPO and
smoothed to satisfy a topographic stiffness-ratio of 0.2
\citep{haidvogel_1999}. The minimum depth used is 3 m. The configuration uses a fourth-order horizontal advection
scheme for tracers, a third-order upwind advection scheme for momentum,
and the turbulence closure scheme for vertical mixing by
\cite{mellor_1974}.

The simulations were initiated from November 2003 and ran until the end of 2007 with
realistic initial and boundary conditions from the global model HYCOM, thus providing daily outputs
over years 2003-2007, overlapping with the timing of the field survey.
HYCOM \citep{hycom_2009} has horizontal resolution of 1/12$^o$ and the data is available
at 33 vertical layers. This model assimilates observations from
several sources including satellite altimetry, satellite and \emph{in situ} temperatures, and vertical temperature and salinity profiles
from XBTs and ARGO buoys. The offline nesting procedure employed
here used a nudging region of 40 km along the model
boundaries. In this layer, the 3d model variables (temperature,
salinity and currents) were pushed towards HYCOM outputs with a
time scale of 8h. The nudging was set to a maximum at the boundaries,
decaying sinusoidally to zero inside the nudging layer.
At the boundaries, radiation conditions
were used for the baroclinic variables \citep{marchesiello_2001}. Sea surface height and barotropic
currents from the parent models were imposed at the boundaries following
\cite{chapman_1985} and \cite{flather_1976}.
\added{Tidal forcing was employed at the lateral boundaries, as amplitudes and phases of tidal elevation and as barotropic tidal ellipses of the main diurnal and semi-diurnal tidal constituents. This forcing was obtained from the 1/12\textdegree Atlantic dataset of TPXO
\citep{egbert2002}.}

ROMS has been applied in many coastal modeling studies in the region such as studies
of dispersion and recruitment of larvae \citep{mma_cjfas_2008}, river plumes
\citep{Otero2008} and pollution transport \citep{mma_mpb_2013}. While these studies
were based on climatological open boundary conditions, offline nesting has been adopted
in the current work to improve realism. It has been shown that using an assimilating
parent model improves the ROMS skill \citep{hetland_2006}, even when parent
and child are forced with different atmospheric data \citep{mma_jgr_2013}.
Other works have used the same nesting procedure and HYCOM as parent model
\citep[e.g.][]{barth_2008, zhang_jgr_2012,fennel_jgr_2013,fabiola_csr_2013}.

Surface heat and freshwater fluxes from ERA-INTERIM
\citep{interim_2011} were used, with a resolution of 0.75$^o$.
As wind forcing, data from the Cross-Calibrated Multi-Platform (CCMP)
Ocean Surface Wind Vector Analyses \citep{ccmp_2011} were used. CCMP uses a variational
analysis method to combine data from satellite sources, producing 0.25$^o$
gridded winds every 6 h.

River discharges from the main Portuguese and Spanish rivers were obtained from the
national institutes and were included here by using the procedures described by \citep{Otero2010}.
Flow data of large rivers (e.g. Douro and Minho) come from measurements obtained during the
period of study while others are monthly climatologies (due to absence of data). River temperatures were
set equal to the local climatological air temperatures \citep{dasilva_etal_1994}.

\subsection{Lagrangian Coherent Structures and Finite-size Lyapunov exponents}

We characterize the filamental structure based on the computation of
Lagrangian Coherent Structures (LCS). These are structures that separate different dynamical
regions of the flow, serving as proxies for barriers and avenues to transport
or eddy boundaries \citep{Boffetta2001, HallerYuan2000, dOvidio2004}.
We are interested in a fully 3d characterization of the LCS that
delimit the frontiers of the filament. For this we follow work done in
\cite{Bettencourt2012} in
another upwelling region (Benguela) to unveil the dynamics of a mesoscale eddy moving offshore. Thus the LCS are computed as the ridges with high values of the FSLE
fields \citep{dOvidio2004, dOvidio2009, Bettencourt2012,
	Bettencourt2013}. It is important to mention that the particle trajectories
are integrated
backwards-in-time so that we compute the so-called attracting LCS. These are the
most relevant for their physical implications since particles typically approach and
spread along them \citep{dOvidio2004,Bettencourt2012}.
\added{LCS are tolerant to imprecisions in the velocity field \citep{Ismael2011}. The time scales for all the bio-physical processes that we study (for example, a significant variation of the position of the eddies) are much larger than one day, which makes it acceptable to use daily model outputs.}

The FSLEs, $\lambda$, measure  particle dispersion at finite scales and were
introduced to study non-asymptotic dispersion processes such as stretching at finite
scales and bounded domains \citep{Artale1997, Aurell1997,Boffetta2001}.
They are  defined as:
\begin{equation}\label{FSLE_1}
\lambda(d_{0},d_{f};\mathbf{x}_0,t)=\frac{1}{\tau}\log\frac{d_{f}}{d_{0}} \ ,
\end{equation}
where $\tau$ is the time it takes for the separation between
two particles, initially $d_{0}$, to reach a value $d_{f}$.
In addition to the dependence on the values of $d_0$ and $d_f$,
$\lambda$ depends also on the initial position of the particle $\mathbf{x}_0$
and the time of deployment $t$.

The $\lambda$ fields in the region of interest (Fig. \ref{fig:west-iberia}) were computed following the method
of \cite{Bettencourt2012}, using $d_0$ equivalent to one half of the
horizontal spacing of the ROMS grid.
In brief, the FSLE is computed in a 3-d grid of initial conditions, composed of constant depth layers, in a quasi-3d fashion: the particle trajectories are computed using the 3-d velocity field but the calculation of the interparticle distance  only considers horizontal distances. In addition,
the ridges delimiting the filament are computed in 3d space, following the methodology
outlined in \cite{Bettencourt2012}. To do this, the $\lambda$ field was smoothed \citep{Garcia2010}
and high-pass filtered to remove the small-scale noise. The smoothing process has the side-effect of smearing the FSLE gradients, which reduces the ridge strength and \replaced{make}{makes} it harder for the extraction algorithm to delineate the complete filament boundaries.

The vertical resolution of the FSLE computation is variable with 30 horizontal levels between 5
and 200 m depth, clustered near the surface. The final distance threshold $d_f$ was set at 100 km
and the particles were integrated for 90 days with a 4th--order Runge-Kutta method,
using the ROMS velocity fields from June to December 2007. We note that with a 90 day integration period
the \replaced{smalles}{smallest} $\lambda$ that can be measured is $0.05\:day^{-1}$ , but for plotting purposes,
we assign a value $\lambda=0$ to the initial conditions for which the final separation
$d_f$ is not reached after the 90-days integration period.
FSLE fields were computed daily from late August to early October
when filament activity is stronger \citep{Haynes1993}.

Offshore transport through the filament was further investigated by computing particle trajectories
released at the filament root. Particles were released at 9\textdegree 30'W on the
26\textsuperscript{th} of September 2007 at 7, 12, 24, 29 and 34 m depth. At each depth, seven particles
were released along a line between 41\textdegree 30'N and 41\textdegree 35'N, inside the
filament root. Particle trajectories were integrated until the year's end (i.e. during about 3 months). Only the particles that travelled
offshore of 10\textdegree 30'W were considered in the analysis. Temperature and salinity fields were interpolated along the particle trajectories \citep{chenillat2015quantifying}.


\section{Results and discussion}\label{sec:c6-results}

\subsection{Oceanographic background and mesoscale structures from FSLE fields}
\label{subsec:c6-3d-fsle}

\begin{figure}
	\centering
    \includegraphics[width=\textwidth]{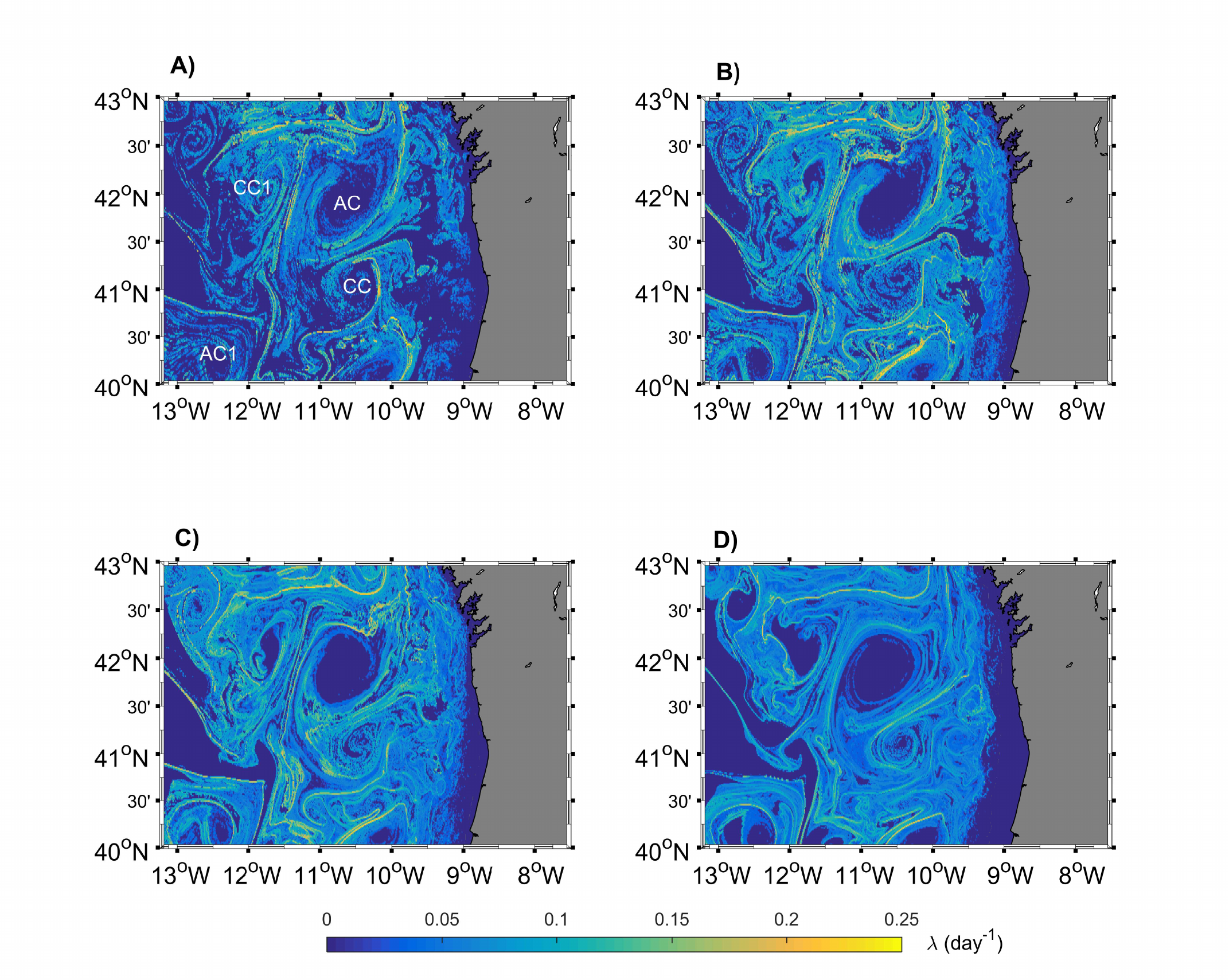}
	\caption{Backward FSLE field on 26 September 2007.
		A) 9.8 m depth;
		B) 17.1 m depth;
		C) 31.8 m depth;
		D) 96.8 m depth. For the labels in panel A) see text in section \ref{subsec:c6-3d-fsle}.}
	\label{fig:backward-FSLE}
\end{figure}

The dynamics in the region is dominated by a complex pattern of mesoscale and submesoscale
features at different depths as shown in Figure \ref{fig:backward-FSLE}.
Lines of high $\lambda$ are found mainly on the boundaries of the dominant mesoscale structures.
The surface field is markedly noisy, while the deeper levels show less small-scale features.
The coastal strip, of about 20-50 km width (up to around 9\textdegree 30'W),
is characterized by weak $\lambda$ and patches where $\lambda<0.05\:day^{-1}$, identifying zones of
weak dispersion of fluid elements.

\begin{figure}
	\centering
	\includegraphics[width=\textwidth]{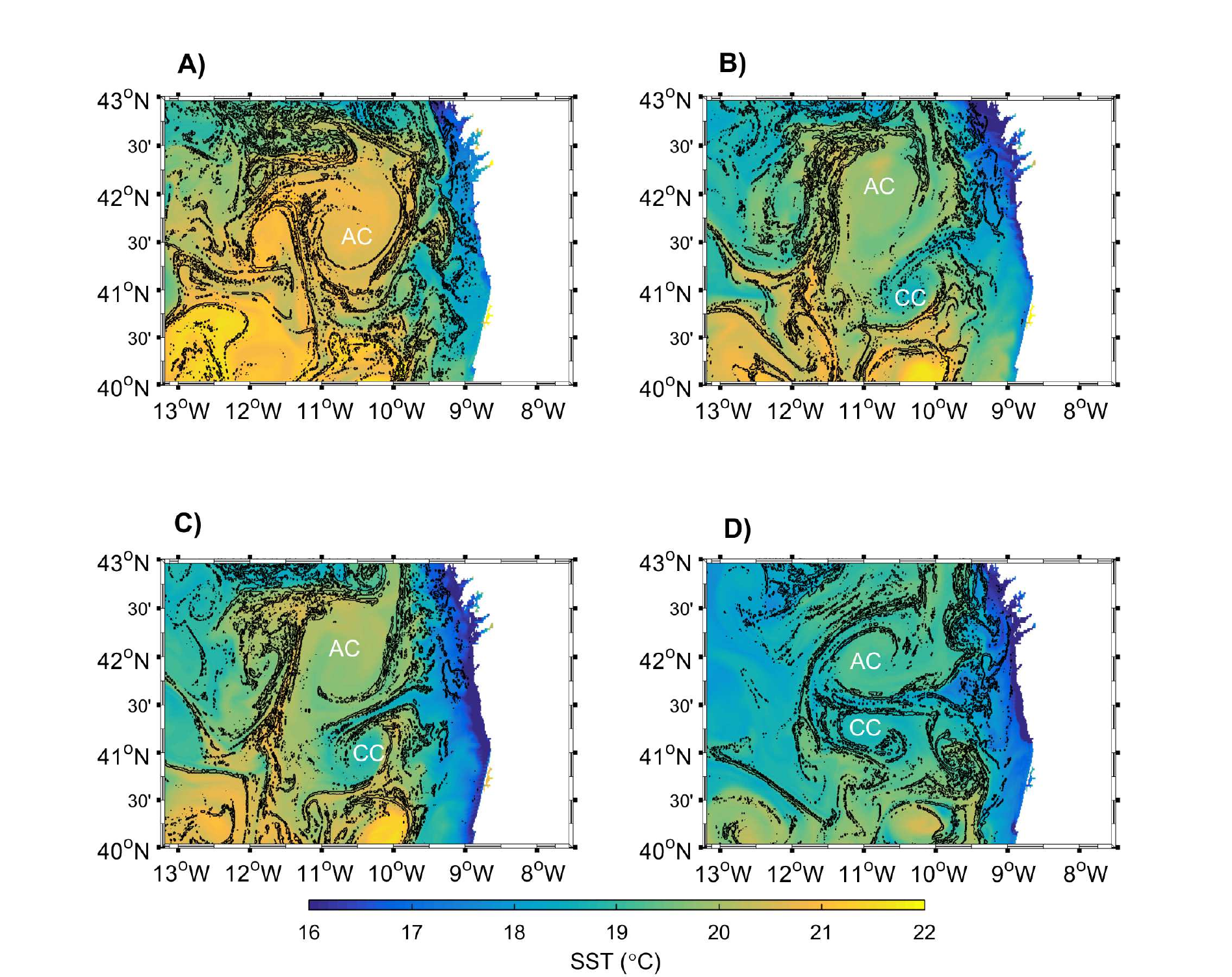}
	\caption{Filament evolution. A) SST and backward FSLE contours (0.1 day\textsuperscript{-1}) on
		A) 09/09/2007;
		B) 20/09/2007;
		C) 24/09/2007;
		D) 06/10/2007. For the labels see text in section \ref{subsec:c6-3d-fsle}.}
	\label{fig:filament-evol}
\end{figure}

On September \replaced{24}{26}, 2007, the dominant structure is a large anticyclonic eddy with centre located at 42\textdegree N / 10\textdegree 30'W
and a diameter of $\sim$100 km (labelled AC in Figure \ref{fig:backward-FSLE}A and Figure \ref{fig:filament-evol}).
Westward of this
warm-core eddy, a cyclonic eddy (centered at 42\textdegree 15'N / 11\textdegree 45'W) is found,
labelled CC1 in Figure \ref{fig:backward-FSLE}A
which, together with the former eddy, assemble a mushroom-like dipole structure. Between them, an elongated
high-strain region can be found, signalled by the concentration of parallel high $\lambda$ lines.
The dipole ispulling fluid from the south along its meridional axis.
South of the latter cyclonic eddy we find an additional anticyclonic eddy (centred at about
40\textdegree 15'N / 12\textdegree 30'W), labelled AC1 in Figure \ref{fig:backward-FSLE}A forming another dipolar structure with the CC1 eddy, a situation previously observed \citep{Fedorov1989}. This second
dipole has an axis orientated along the zonal direction and is pulling fluid towards the shelf.
The fluid advected zonally by this second dipole is slightly colder (see Figure \ref{fig:filament-evol}C)
than the water within the core of the AC1 eddy. It thus appears as a colder water tongue along the dipole axis,
that is deflected southward due to the AC1/CC eddies.

Concerning the coastal patches of small $\lambda$, the one centred on
41\textdegree 30'N / 9\textdegree 30'W at 9.8 m depth (Figure \ref{fig:backward-FSLE}A) has a
thin south-westward extension that provides an offshore transport pathway for coastal
waters in-between the AC and CC eddies. This is the filamentary structure that will be thoroughly studied in the rest of the paper. The weak dispersion evidenced by $\lambda<0.05\:day^{-1}$ indicates that the fluid
particles released from this region will evolve, backward in time, along nearby trajectories
with very small divergence between particle pairs. In other words,
nearby particles passing through these regions come from nearby initial positions.
Observing the deeper levels in Figure \ref{fig:backward-FSLE}, we can see that
the $\lambda<0.05\:day^{-1}$ patch is present at 17 m depth (Figure \ref{fig:backward-FSLE}B)
and at 32 m depth (Figure \ref{fig:backward-FSLE}C), becoming smaller with depth.
This could indicate the presence of an upwelling cell, i.e. a region through which subsurface fluid
parcels rise to the surface and subsequently drift offshore, as shown in section \ref{subsec:c6-filament-evol} .

There are other regions where $\lambda$ is small. The AC eddy
shows $\lambda<0.05\:day^{-1}$ in its core at all depths shown in Fig. \ref{fig:backward-FSLE}.
Note however that the AC eddy is not completely enclosed by high $\lambda$ lines since it presents openings with $\lambda \neq 0$ at its north-east and south-west sectors (e.g. Fig. 2B). High $\lambda$ lines symbolize "coherent" boundaries that clearly separate waters in the AC core from the surrounding waters. In contrast, areas with small $\lambda$, bounded by parallel high FSLE lines, represent preferential avenues of transport for water to enter and exit the AC eddy. They also indicate the likely pathways of transport: water of northern origins enter the AC through the north-east open boundary; coastal waters that were previously advected offshore through the filament (in between AC and CC) then tend to recirculate into AC via the south-west open frontier.These pathways are imposed by the anticyclonic circulation and have been confirmed by particle trajectories analysis.


\subsection{Formation and evolution of the coastal filament}
\label{subsec:c6-filament-evol}
The rest of the paper will be devoted to the detailed study of the filamental structure identified in the previous section flowing (south)westwards between eddies AC and CC.
It was formed during the 2\textsuperscript{nd} half of September 2007 and was perfectly
noticeable on September 20 (Figure \ref{fig:filament-evol}B, C) as a cold water tongue connecting the
coastal recently upwelled water with a cyclonic recirculation centred around 41\textdegree N and
10\textdegree 30'W, labelled CC in Figure \ref{fig:filament-evol}B, C and D.

The physical mechanism for the formation of the  filament is beyond the scope of this paper, however we will in this section provide a short description of its formation.
The cold water filament started as a coastal cold water pool that is entrained toward warmer waters
offshore of the upwelling front. This pool is centred at 41\textdegree 15'N and 9\textdegree 30'W on
September 9 (Figure \ref{fig:filament-evol}A). Adjacent to this cold water intrusion, the cyclonic
recirculation still in an initial stage is located slightly to the south-west.

The filament formation could result from the interaction between the anticyclonic eddy AC
and the coastal topography.
Indeed, several observational and numerical studies documented the association
of mushroom-like dipolar structure with the thin filaments favouring cross-shore water
transport \citep{Churchill1986,Oey2004,Sutyrin2010,Meunier2010,Rossi2013}. In particular,
\citet{Sutyrin2010} and \citet{Meunier2010} used numerical simulations to study how along-shore
currents interact with the shelf bathymetry to create topographic eddies. They showed that an anticyclonic
eddy first develops upstream and it is closely followed by the formation of a cyclonic secondary
circulation downstream. Fed by shelf waters with high potential vorticity, both anticyclonic and cyclonic
vortices interact as a dipolar structure which forms an elongated, trapped and narrow filament along its axis.

The dynamics of the filament seem tightly linked to the evolution of both anticyclonic (AC)
and cyclonic (CC) eddies forming the dipole. On September 20th, the CC eddy has grown and the
potential temperature
(henceforth temperature) is lower than the surrounding waters due to the injection
of recently upwelled waters within the cold water filament. Four days later (Figure \ref{fig:filament-evol}C) the
filament axis is mainly zonal and its southern seam continues exporting colder water offshore.

Remarkably, there is a parallel offshore motion of warmer waters in the northern flank of the cold water filament.
The northern warm water channel, influenced by the adjacent AC
eddy, pulls water of about $\theta \simeq 19-20$\textdegree C that is isolated from the recently upwelled
waters due to a series of transport barriers (signalled by the high FSLE lines along the contact zone
between the eddy and the upwelling front, Figure \ref{fig:backward-FSLE}A, B). The cold water filament
starts on the shoreward side of the upwelling front, thus pulling
colder water of about $\theta \simeq 17-19$\textdegree C offshore (Figure \ref{fig:filament-evol}C).

The filament continues developing. On October 6 (Figure \ref{fig:filament-evol}D), the mushroom-like dipole
formed by the CC and AC eddies (now exhibiting almost similar diameters of 80 km) is well
developed and drifts offshore. It results in an extension of the filament length which continues transporting
colder water offshore. Note that the temperature differences between the filament core and its surroundings
have diminished as compared to 2 weeks before which could be related to a new upwelling pulse and/or to some
air-sea processes (such as solar warming or wind-induced cooling)
have diminished relatively to the previous weeks as expected due to the onset of Autumn conditions (lower solar radiation and northern wind associated with colder air temperature). The filament continues to evolve
further but its signature on SST is significantly reduced.

On the 24\textsuperscript{th} of September 2007, which is the date were the
FSLE fields most clearly represent its structure (Figure \ref{fig:filament-evol}C),
the filament under study is about 140 km long and 10 km wide. The
northern warmer channel has a width of 15 km.
At it's maximum extension,
the filament is about 230 km long. Those dimensions are
consistent with observations \citep[e.g.][and references therein]{Rossi2013}.

\subsection{Vertical properties of the filament}
\label{subsec:c6-filament-lst}
\begin{figure}
	\centering
    \includegraphics[width=\textwidth]{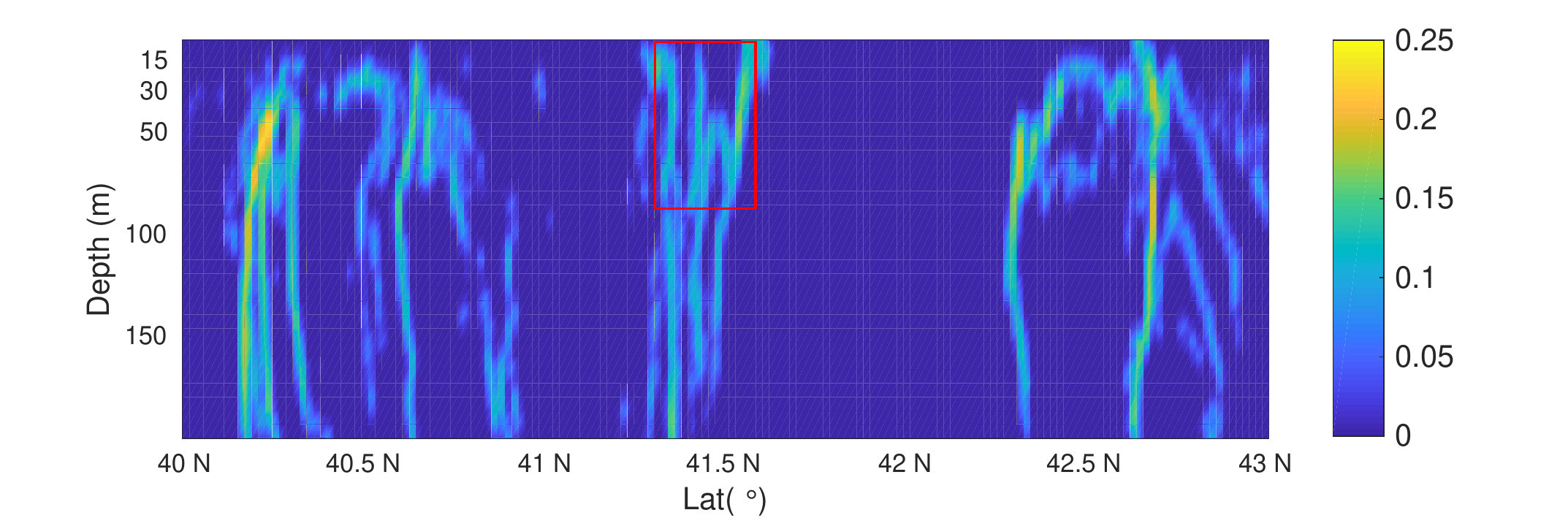}
	\caption{Vertical section of backward FSLE taken at 10\textdegree 15' W on
		26/09/2007. The FSLE field was smoothed and filtered to improve clarity and highlight the main features of the FSLE field. The
		filament area is highlighted in the red rectangle.}
	\label{fig:vertical-section-bFSLE}
\end{figure}

The filament appears as a thin elongated region of $\lambda<0.05\:day^{-1}$ at 9.8 m and 17.1 m depth (Figure
\ref{fig:backward-FSLE}A, B). The signature is absent on deeper levels, indicating
the shallow nature of this corridor transporting shelf water to the offshore. A vertical
section through the backward $\lambda$ field on the \replaced{24}{26}\textsuperscript{th} of September 2007
(Figure \ref{fig:vertical-section-bFSLE}) shows its vertical extent to be about 30 m deep, in good
agreement with previous observations-based estimates \citep{Rossi2013}.

At the northern border of the filament, there is a shallow strip that constitutes the northern channel
through which relatively warmer waters recirculate. The boundaries of the AC eddy intersect
the plane northward of the filament, indicating that the surface section of the eddy is slanted toward the north with respect to the deeper levels,
especially its northern flank, also characterized by the highest $\lambda$. The CC eddy is seen
south of the filament, presenting higher values of $\lambda$ in its core, centred at
41\textdegree N.
A possible reason for the difference in the $ \lambda $ field inside both eddies is that, because the CC eddy is young compared to the AC eddy, the backward $ \lambda $ field in its interior reflects the different origins of the waters that formed the eddy, while the waters in the older AC eddy have been isolated from the exterior for more than 90 days (the FSLE integration time).

\subsubsection{Thermohaline vertical structure}

\begin{figure}
	\centering
    \includegraphics[width=0.7\textwidth]{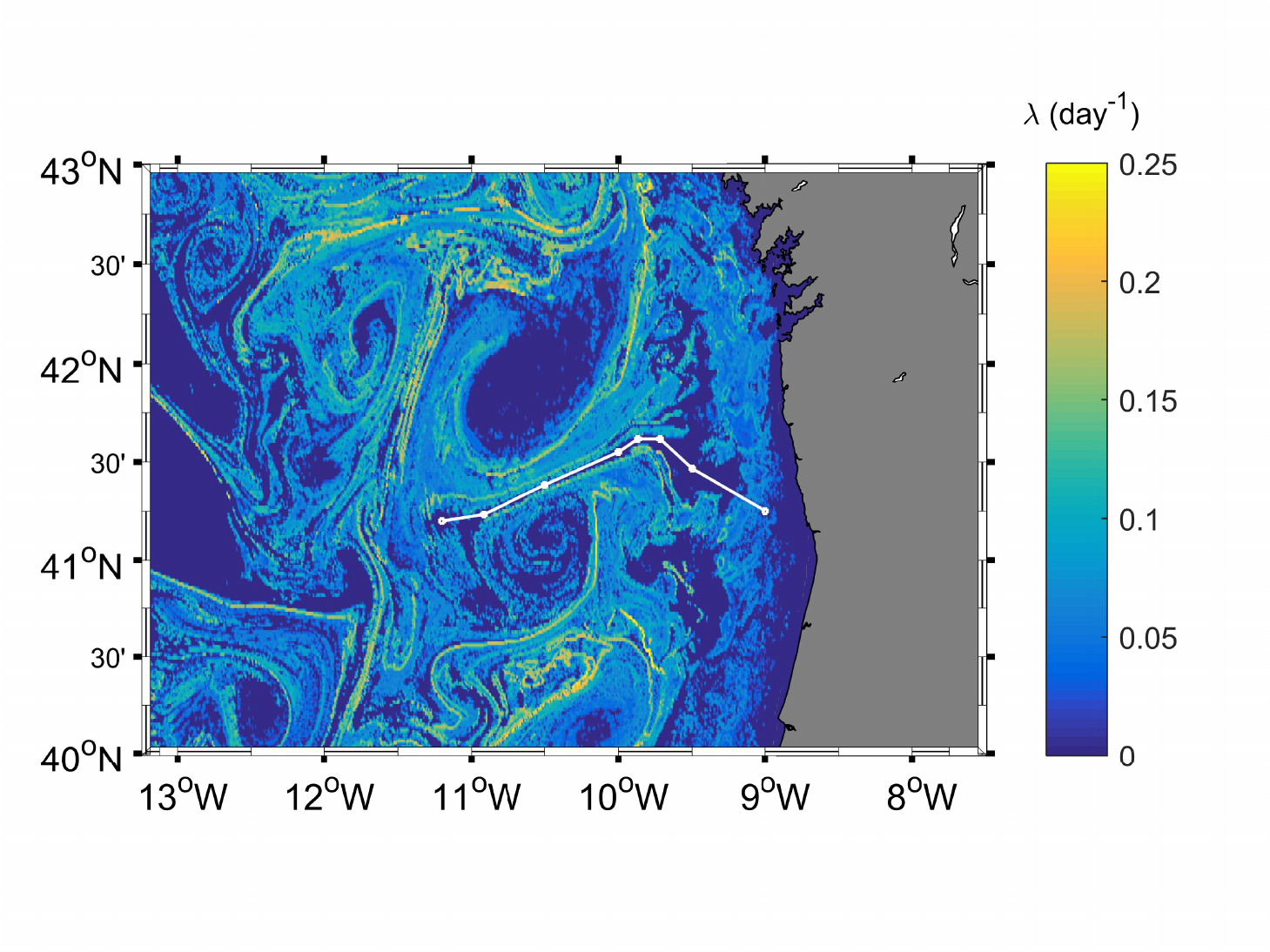}
	\caption{Location of the profile path (white line) taken along the filament on the 26/09/2007. FSLE map at
		14.7 m depth.}
	\label{fig:filament-profile}
\end{figure}

\begin{figure}
	\centering
    \includegraphics[width=\textwidth]{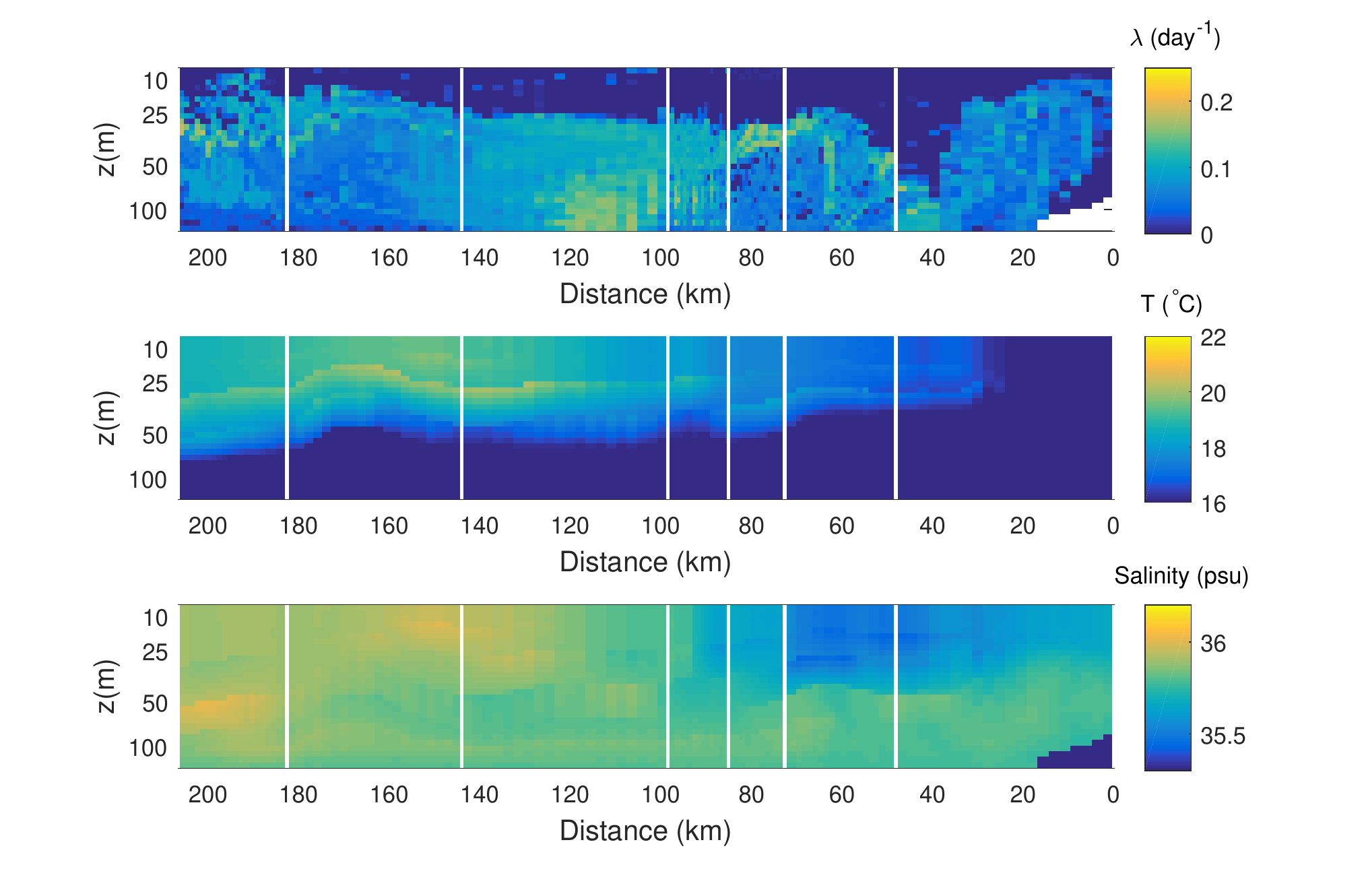}
	\caption{Profiles along filament on 26/09/2007. A) Backward FSLE; B) Temperature; C) Salinity. White vertical
		lines represent locations of profile of the white circles marked along the profile (white line of Fig. \ref{fig:filament-profile}). Fields were interpolated linearly at 20 equally spaced points in between nodes.
		Distance along profile increases towards the offshore.}
	\label{fig:profiles-filament}
\end{figure}

$\lambda$, $\theta$ and salinity fields were interpolated linearly along a selected section within
the filament (white line on Figure \ref{fig:filament-profile}) on September 26.
The cross-shore and vertical variations of these quantities are shown in Figure \ref{fig:profiles-filament}.
The filament appears as a shallow feature in the $\lambda$ field (Fig. \ref{fig:profiles-filament}A)
with central portions not deeper than 25 m. At the root, approximately 40 km from the coast, the filament is deeper than 50 m but \replaced{it's}{its} size diminishes with increasing depth. This can also be observed in Figure \ref{fig:backward-FSLE}, where the size of the
$\lambda<0.05\:day^{-1}$ area at the root of the filament decreases with depth. Offshore, the filament becomes shallower until it vanishes within open ocean waters
at $\sim$ 200 km offshore. We can observe that before reaching the offshore tip of the filament, the $\lambda<0.05\:day^{-1}$ region
has curved southward to join the CC eddy that is receiving the water exported through the filament.
The temperature profile (Fig. \ref{fig:profiles-filament}B) indicates that the filament
is composed of recently upwelled water ($\theta \leq 17$\textdegree C) at the shoreward tip that progressively warms
up as
it flows through the filament. There is a noticeable vertical $\theta$ gradient located at about 30-60 m below the surface.
The salinity  profile (Fig. \ref{fig:profiles-filament}C) shows a pool of freshwater located
at the root of the filament down to 30 m. This type of low salinity lens is commonly observed
in the region \citep{Otero2008,Otero2010,Rossi2013} but it is still unclear if it originates from
continental freshwater inputs and/or from recently upwelled waters (such as the low salinity Eastern
North Atlantic Central Waters). The salinity then gradually increases westward until reaching a maximum
140 km offshore from the filament root. The thermohaline structure of this simulated filament appears to be in very good agreement with the observations of \emph{Rossi et al. (2013a).} In particular, both vertical and horizontal structures as well as gradients are very similar when comparing our Fig. 6B, C and Fig. 7b, c of Rossi et al. (2013a).

Note that between 140 and 180 km from the coast, there is a clear uplifting of isotherms, also noticeable as a doming
in the $\lambda$ field (Fig. \ref{fig:profiles-filament}A), caused by the cyclonic eddy situated
to the south of the profile. This is a typical situation where small structures superimpose
their signatures on top of the larger (mesoscale) filament \citep{Rossi2013JGR}.

\subsubsection{Particle trajectories within the filament}
\label{subsub:particle_trajectories}

\begin{figure}
	\centering
    \includegraphics[width=0.8\textwidth]{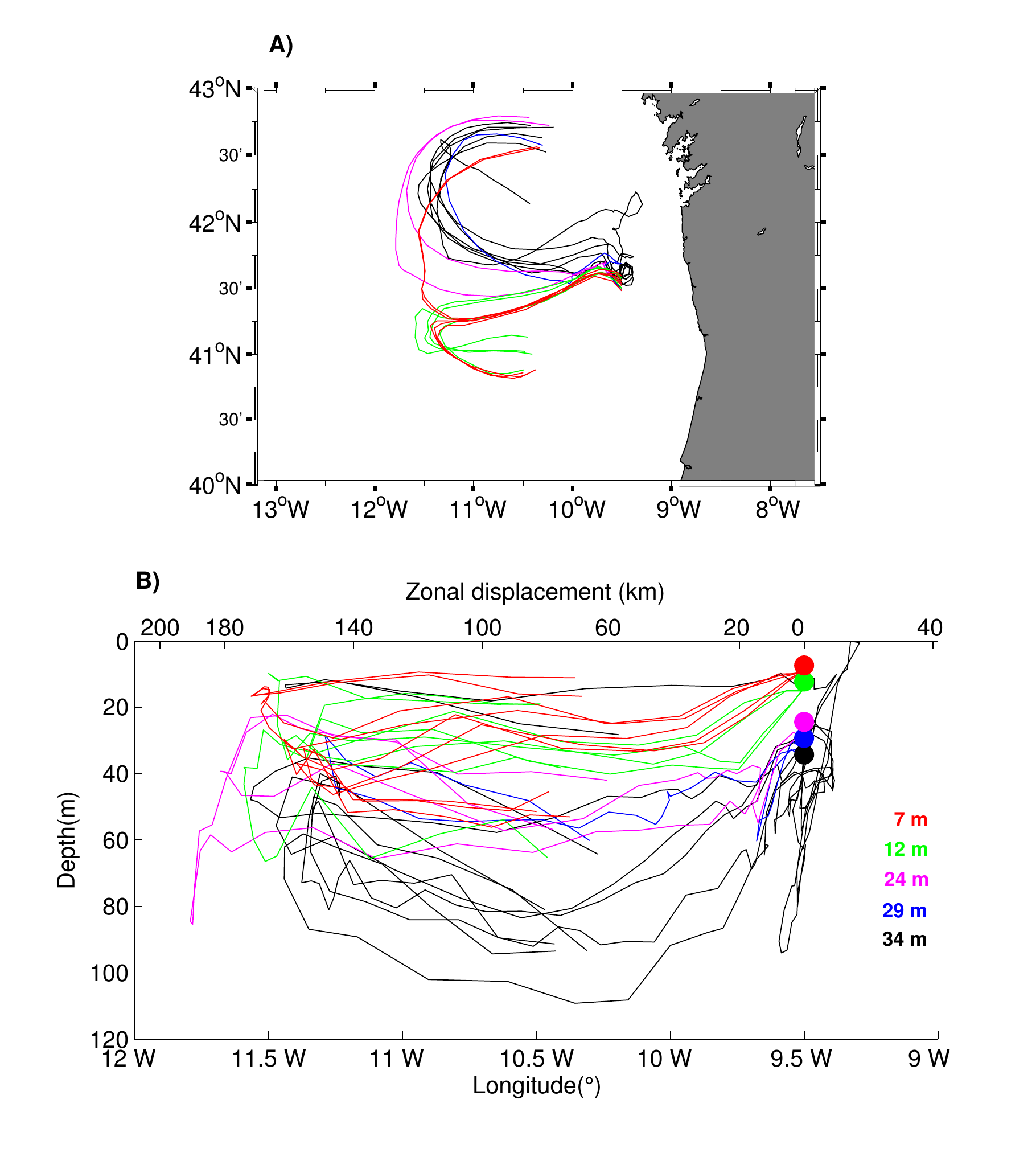}
	\caption{Trajectories of particles released at the root of the filament. A) Horizontal projection of the trajectories;
		B) Vertical and longitudinal displacement along the particles trajectories. Particles were released at 9\textdegree 30' W at
		26/09/2007 at 7, 12, 24, 29 and 34 m depth and were integrated until year's end.
		Only the particles that traveled offshore of 10\textdegree 30' W are shown, until
		they crossed the same longitude in the opposite direction.}
	\label{fig:particles-trajectories}
\end{figure}

Tight clusters of five particles were released from five initial depths (Figure \ref{fig:particles-trajectories}) to illustrate the dynamics of the trajectories starting at the root of the filament.
The horizontal particle trajectories (Figure \ref{fig:particles-trajectories}A) clearly separate
the set in two groups depending on the release depth: shallower
released (7 and 12 m) particles flow through the filament and then into the cyclonic
circulation while deeper released particles turn northward into the anticyclonic eddy
when they exit the filament. That the release depth is the factor determining the fate of the particles should not be a surprise\deleted{d} since the filament boundaries are slanted in the vertical and deeper particles will be on one side of the boundary while shallower particles will be on the other side.
Note some exceptions occur such as some particles released
at 7 m that follow the filament but end-up circulating around the anticyclonic eddy or the
particle released at 34 m that travels northward before entering the anticyclone.

The maximum zonal displacement is $\sim$ 200 km (Figure \ref{fig:particles-trajectories}B). The shallow particles that flow through the filament
experience a subduction and reach their deepest positions $\sim$ 60 km after
release. They then begin to rise briefly until the 100 km offshore mark. The "red" particles
enter the cyclonic circulation and sink, while the "green" particles remain approximately
at the same depth. The two red particles that enter the anticyclonic circulation eventually
rise when their trajectories curve to the north.
The different behaviour of the "green" and "red" particles is most likely caused by the fact that red particles, circulating further from the cyclone's center, tend to follow deeper isotherms than the green particles.

\begin{figure}
	\centering
    \includegraphics[width=\textwidth]{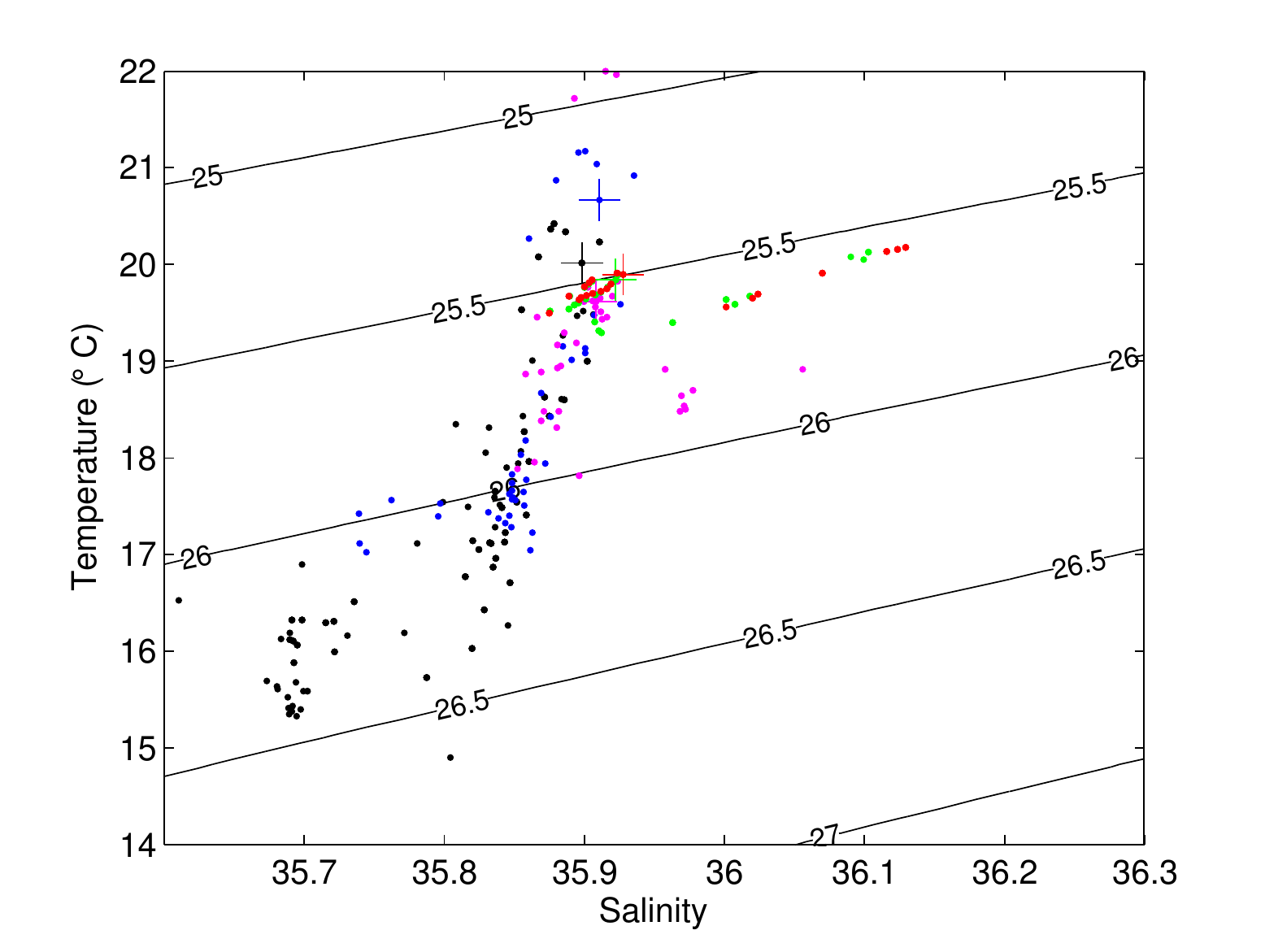}
	\caption{Temperature-Salinity ($\theta$-S) diagram of particles released at the root of the filament. Black lines
		in the diagram are constant density lines in $kg/m^3$. Crosses indicate the $\theta$-S properties at the
		release location.}
	\label{fig:particles-TS}
\end{figure}

The temperature-salinity ($\theta$-S) diagram (Fig. \ref{fig:particles-TS}) associated with these particles
reveals that their initial properties (at the release depth) are quite similar in terms of
salinity but distinct in terms of $\theta$, with variations as much as 1\textdegree C
(crosses on Figure \ref{fig:particles-TS}).
The coldest water is found at 24 m depth and the warmest at 29 m depth, warmer than the
shallow release locations. This is caused by the location of the sharp $\theta$ gradient at
the bottom of the mixed layer in the filament root. The temperature profile
(Fig. \ref{fig:profiles-filament}B)) shows that this gradient coincides with
the two deeper sets of particles and that these subsurface waters can be slightly
warmer than surface layers.
The two shallower sets (red and green particles) then essentially evolve along pathways during which their density
is conserved (Fig. \ref{fig:particles-TS}), representing transport along isopycnals. The remaining sets (blue, magenta and black) show contrasting
evolutions with important changes of their temperatures, suggesting the presence of diapycnal mixing (i.e. transport of water parcels across isopycnals, resulting in changing thermohaline properties).

\subsubsection{Horizontal gradients of temperature in the filament}
\label{subsubsec:c6-3d-lcs-ts-grads}

\begin{figure}
	\centering
    \includegraphics[width=\textwidth]{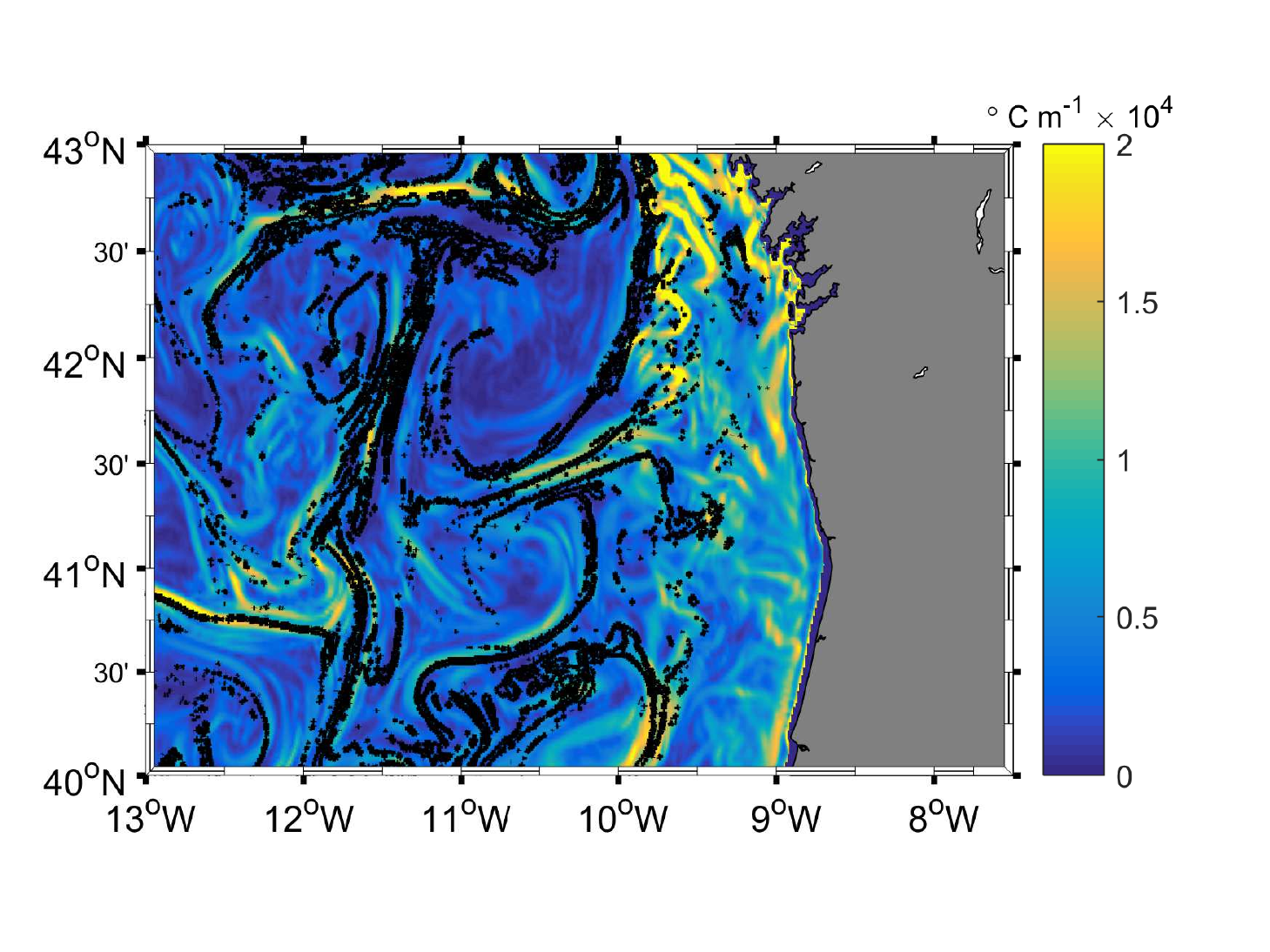}
	\caption{Map of the norm of the horizontal temperature gradient at 12.25 m depth,
		together with contours (black lines) of backward FSLE at the 0.12 day$^{-1}$ level on 26
		September 2007.}
	\label{fig:2d-Tgrad-fsle}
\end{figure}

Next, we analyse the norm of the horizontal gradient of temperature on September \replaced{24}{26}, 2007
at 12.25 m depth together with contours of $\lambda$ (Fig. \ref{fig:2d-Tgrad-fsle}). The filament
boundaries appear both as high $\lambda$ regions and as strong temperature gradient norm. This coincidence
is verified in several other regions of the domain, but not exactly everywhere.

We note specially the strong $\theta$ gradients along $\sim$ 9.5$^\circ$W that signal the shelf break
front
separating cold recently upwelled water masses from warmer offshore waters. This is not a continuous
feature but it presents instead a complex wave-like aspect with the dominance of submesoscale
patterns. It suggests that vertical dynamics becomes more important closer to the coast. This could explain
the poor matching between $\lambda$ maxima and strong $\theta$ gradient features in the shelf region since the
$\lambda$ computation was tuned to mesoscale features.

On the other hand, the filament boundaries are properly captured by the $\lambda$
and by the $\theta$ gradient fields. The root of the filament appears to the southern
end of a cold water region situated inshore of the frontal features between 41.30$^\circ$N and 42.30$^\circ$N.
Capturing the filament boundaries using both methods returned good agreement probably because the dynamics giving
rise to the filament, the pair of counter-rotating eddies, occurs at the mesoscale and off the shelf.
Also, the fact that particles flowing through the filament conserve their (potential) density (section \replaced{\ref{subsec:c6-filament-lst}}{\ref{subsub:particle_trajectories}})
further supports the finding of matching $\lambda$ maxima and strong horizontal gradient norms. Moreover, through a careful
examination of Fig. \ref{fig:2d-Tgrad-fsle}, it is evident that $\lambda$ maxima match strong $\theta$ gradient
features along the boundaries of most mesoscale eddies and filamentary structures.


\subsection{3d structure of the filament}
\label{subsec:c6-3d-lcs}

The 3d Lagrangian structure of the filament is investigated by extracting ridges in the
3d backward $\lambda$ field. These ridges approximately coincide with attracting Lagrangian Coherent
Structures (LCS) of the flow \citep{dOvidio2004, Mancho2006b,Prantsreview,haller2015lagrangian}. To avoid unnecessary computational effort, the ridge extraction
was limited to a subregion that contains the filament, i.e. between 11\textdegree 20'W
and 9\textdegree 30'W, 41\textdegree N and 41\textdegree 45'N and over 7-70 m (see
Fig. \ref{fig:west-iberia}).

\begin{figure}
	\centering
    \includegraphics[width=\textwidth]{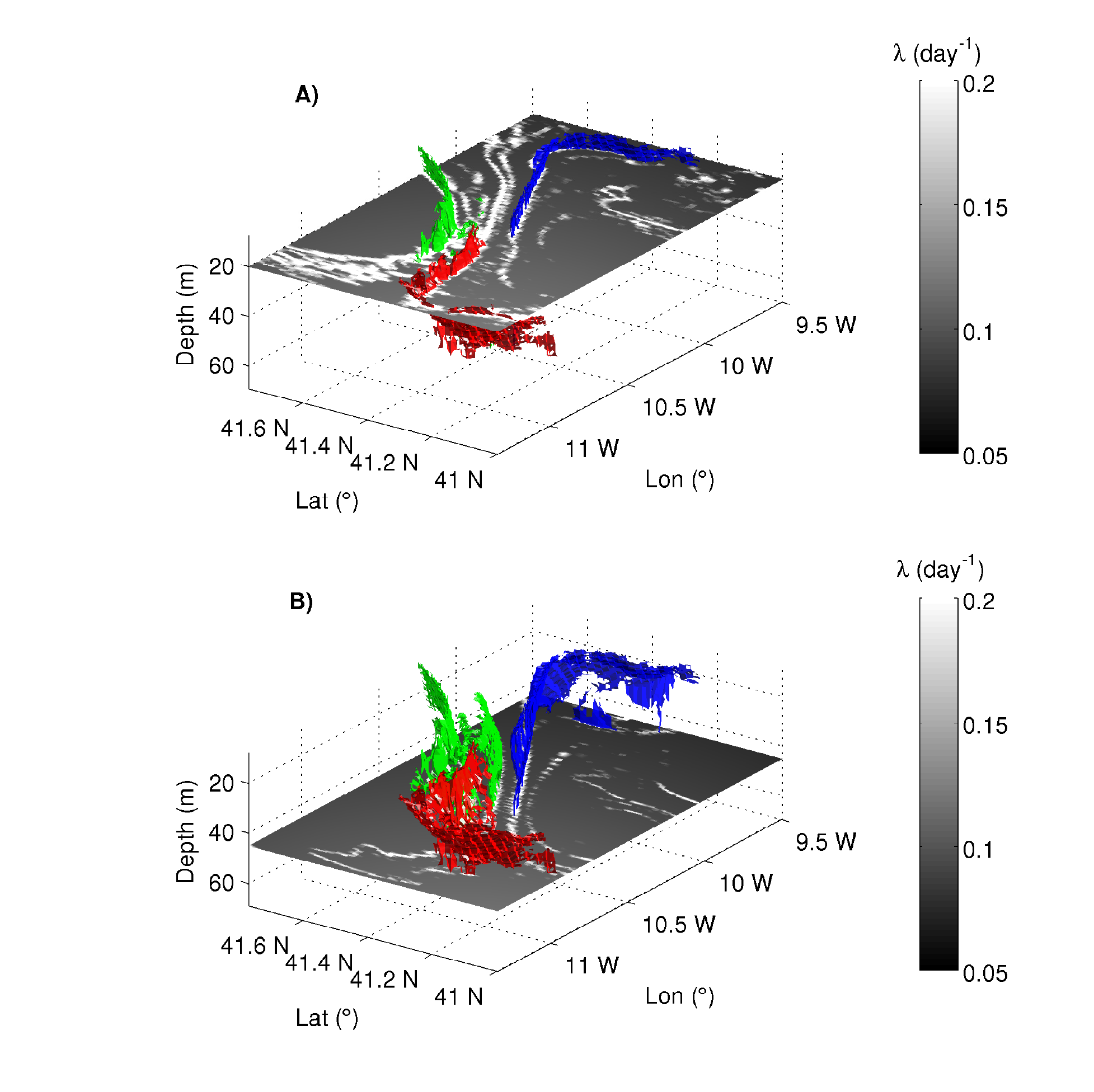}
	\caption{Two views of the ridges of the 3d FSLE field defining the filament extracted on 26/09/2007.
		Green: boundary of the anticyclone; Blue: boundary of the cyclone; Red: Ridge that separates the cold water filament from the northern warm water channel. A) FSLE field at 20 m depth;
		B) FSLE field at 45 m depth.}
	\label{fig:3d-lcs}
\end{figure}

The relevant ridges on September\deleted{,} 26\textsuperscript{th}\added{,} 2007, that best reveal the Lagrangian
dynamics of the filament, are shown in Figure \ref{fig:3d-lcs}. They are the boundary of the
anticyclone (in green), the boundary of the cyclone (in blue) and the LCS that separates the cold water filament from the northern warm water channel (in red).
Levels of the $\lambda$ field are plotted together with the 3d ridges at 20 m depth (Fig. \ref{fig:3d-lcs}A)
and at 45 m (Fig. \ref{fig:3d-lcs}B). As discussed by \cite{Bettencourt2012}, the ridges fall
on the high $\lambda$ lines which gives confidence to our 3d computation.
Fluid elements flow along attracting ridges of the FSLE field, so the blue ridge is moving fluid offshore
from the filament root. The green ridge, on the other hand is separating the fluid inside the anticyclone
from the water that is outside but flows around it. Note that, at least at the surface, this water mass is
warmer than in the filament core (Fig. \ref{fig:filament-evol}C) so there should be a barrier separating
these two streams and preventing them from mixing. On the $\lambda$ map at 20 m depth
(Fig. \ref{fig:3d-lcs}A), there is a line of high $\lambda$ separating these regions (white
lines), but the automatic extraction process found no ridge there. In contrast, it found a ridge separating
both streams at 45 m (Fig. \ref{fig:3d-lcs}B). It is possible that the ridge extraction threshold was set higher than the intensity of this high
FSLE line or that the smoothing of the FSLE field reduced artificially the ridge strength, thereby rendering it invisible to the extraction algorithm.
However, the purpose of this analysis is the \replaced{deliniation}{delineation} of the
filamentary structure, so that the threshold was set accordingly after preliminary tests (not shown).
The retained threshold is a compromise between a value small enough to extract the
relevant boundaries but big enough not to capture the many smaller-scale less energetic structures.

\begin{figure}
	\centering
    \includegraphics[width=\textwidth]{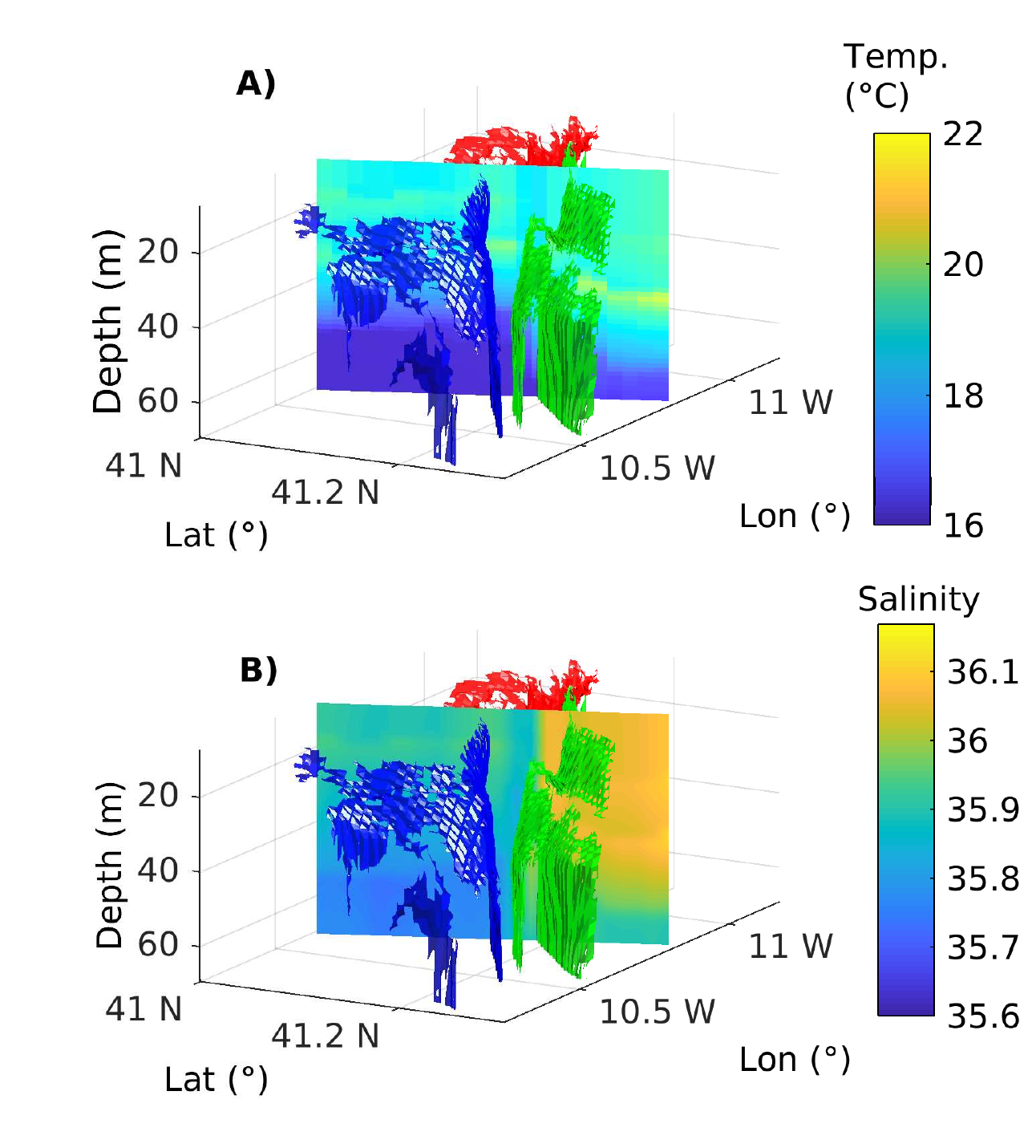}
	\caption{Ridges of the 3d FSLE field defining the filament extracted on 26/09/2007.
		Green: boundary of the anticyclone; Blue: boundary of the cyclone; Red: Ridge that separates the cold water filament from the northern warm water channel. A) Temperature.
		B) Salinity. Temperature and salinity are shown in a vertical plane normal to the filament
		axis that passes through 10\textdegree 27' W. Note opposite perspective relative to Figure \ref{fig:3d-lcs}.}
	\label{fig:3d-lcs-ts}
\end{figure}

The 3d shapes of the LCSs delimiting the filament, and their relationship with the thermohaline structure, are
clearly depicted in Figure \ref{fig:3d-lcs-ts} from another perspective (looking along the filament axis towards
offshore). The 3d LCSs are located around and
at the transition between both mesoscale structures. The LCS which separates both eddies and also the cold and warm
water offshore flows (red) is connected to the anticyclonic LCS (green, see Figure \ref{fig:3d-lcs}). Although
a discontinuity is visible (related to the ridge extraction process) the green LCS constitutes
a coherent curtain in the vertical separating clearly the filament from the adjacent anticyclonic eddy. The
cyclonic curtain-like LCS (blue) also reveals continuity in the vertical except at the root of the filament
(interrupted between 30 m and 40 m depth). Note also that the \replaced{strength}{size} of the cyclonic LCS (blue) in the
filament root diminishes with depth, confirming the contraction
of the filament root (source region) already documented in Fig. \ref{fig:backward-FSLE}.

As suggested by previous analysis in section \ref{subsubsec:c6-3d-lcs-ts-grads}, the thermohaline
structure of the filament is related to these dynamical barriers. The main temperature and salinity fronts
match the curtain-like 3d LCSs. There are clear temperature differences between surface and sub-thermocline
waters, as well as between the cyclone, where isotherms are raised, and the anticyclone, where they are depressed
(Fig. \ref{fig:3d-lcs-ts}). The cyclone LCS is positioned between the CC \replaced{eddie}{eddy} and the filament, although is does not coincide exactly with the boundary of the cold water due to the warm tongue of water that is being pulled by the CC eddy and visible in Fig. \ref{fig:filament-evol}C. In accord with previous interpretation, the salinity structure (Fig. \ref{fig:3d-lcs-ts} B) reveals
the distinct origins of the anticyclonic and cyclonic branches of the filament: the former has warm,
saltier water while the latter has colder and fresher waters entering the filament from nearshore regions.

\section{Discussion}
\label{sec:discussion}

Coastal/offshore exchanges are the product of mainly two mechanisms: a linear response to
alongshore wind stress in the form of coastal upwelling and non-linear dynamics in the form of mesoscale and submesoscale processes.
\citet{Combes2013} studied the effects of linear (Ekman upwelling) and non-linear (mesoscale eddies)
circulation dynamics on the statistics of advection of coastal waters. They found that the low-frequency
cross-shelf transport of the upwelled water mass is strongly correlated with the pulses of alongshore
wind stress, but that offshore transport of surface coastal properties is modulated by intrinsic mesoscale
eddy activity, in particular cyclonic eddies. Our analyses also revealed that a preferential offshore
transport of coastal waters occured through the shallow filament that fed cyclone CC during its formation.
In particular, the particle release experiment described in section \ref{subsub:particle_trajectories}
showed that particles released closer to the surface were preferentially captured by the cyclone while particles released from deeper locations were captured by the anticyclone, due to the southward slanting of the filament boundaries.
Waters in the filament maintain constant density, suggesting transport along isopycnals, while those below the filament in general cool during their advection offshore, probably due to dyapicnal mixing, although some trace out warming waters.

The CC eddy was fed by the water transported through the southern channel of the filament \replaced{contained}{containing}
the cold coastal upwelled water while the northern channel carried warmer water on the offshore side of the upwelling front in to the AC eddy.
This is similar to the work  in
\citet{Nagai2015}  that showed that filaments are very effective in transporting coastal
material further offshore until they form eddies at their tips which eventually de-structure the
exporting conduit. Cyclonic eddies tend to trap the cold, nutrient, and organic matter-rich waters
of the filaments, whereas the anticyclones formed nearby encapsulate the low nutrient and
low organic matter waters around the filament. Here we did not track the filament for a sufficient period of time to reach this de-structured state but we consistently observed the least coherent boundaries and the smallest temperature gradients at the latest development stage of our filament.


The high--
$\lambda$ lines present in the region allowed us to identify the major barriers and pathways
of fluid transport, organized by intense mesoscale activity
in the form of mesoscale eddies and submesoscale fronts.
The filament itself appears as a thin region of $\lambda<0.05\:day^{-1}$, indicating
that the fluid progresses along the filament, without deformation, towards the offshore.
The FSLE and its ridges appear as suitable tools to study mass export as they allow to delimit
the main exporting structures, at a resolution higher than the underlying velocity field \citep{Ismael2011}.
On the other hand, Eulerian quantities such as passive tracers are more readily available
from gridded datasets or in situ observations but their ability to describe transport structures depends on whether they are conserved or not, as shown in Figure \ref{fig:2d-Tgrad-fsle}.


From the Lagrangian perspective, cross-shelf export appears to be dominated by meso and submesoscale processes as those identified by the extrema of the $ \lambda $ fields. \citet{Nagai2015} showed in the Californian
upwelling that the lateral export of material is largely controlled by mesoscale processes, involving filaments
and westward propagating eddies. While the analysis presented herein is dedicated to a specific filament and therefore does not allow to draw conclusions in a time mean sense, earlier works using the FSLE \citep{hernandez2014reduction} showed that mesoscale activity was responsible for 30--50\% of organic material export in the Benguela upwelling system and \cite{bettencourt2015boundaries} revealed that the stirring of the dissolved O$ _2 $ field by mesoscale Lagrangian structures in the Peruvian upwelling system greatly impacted the oxygen minimun zone there.

For the specific event described in this paper, we were able to define \deleted{the boundaries} the
boundaries of the mesoscale structures that controlled the cross-shelf exchange. The cold water flow was formed between the AC
and CC eddies and the filament itself was connected to the latter as is evident in the $ \lambda $ maps of figures \ref{fig:filament-evol}
and \ref{fig:filament-profile} and in the particle trajectories plot (Figure \ref{fig:particles-trajectories}).
\added{The physical characteristics of the simulated filament studied from a Lagrangian perspective coincide very well with the ones of the filament observed in Rossi et al. [2013a], namely 140 km long, about 10 km wide and up to 30 meters deep. The thermohaline structures of both observed and simulated filaments are also very similar: temperatures smaller or of the order of 17\textdegree C together with a low salinity lens near the base of the filament; when water flows westwards (offshore export) filament waters exhibit warming up to 19\textdegree C and increasing salinity. In both the observed and the modelled filaments strong gradients in water properties separate the interior from the exterior of the structure.}

The filament is limited by the 3d Lagrangian boundaries of the AC and CC eddies, which provide the necessary barriers to mixing of filament waters with outside ones. An additional barrier exists that separates waters that flow around the AC \replaced{eddie}{eddy} from those that feed the CC eddy with cold, upwelled waters.
Note that the different dynamical regimes,
in this case the different fates, of the fluid particles that flow along the filament are reflected in the Lagrangian structures deduced from the FSLE fields.
Overall, our analyses revealed that local mesoscale processes control the offshore transport efficiency of the
upwelling filament. These factors include the origins (e.g. depth) and properties (e.g. is it formed during upwelling pulse or relaxation?)
of the source waters, the presence of mesoscale structures and their characteristics (e.g. their sizes, ages and directions of rotation)
surrounding the filament as well as the internal dynamics within the filament itself (e.g. isopycnal transport or diapycnal mixing).

A novelty of this work is the attempt to provide a 3d characterization of this process by
computing the 3d ridges of the backward $\lambda$ fields. These ridges are proxies to 3d Lagrangian structures that delineate the full 3d cross-shelf exchange pathways.The ridges shown are, after extensive testing, those with a best quality/extension trade-off. This trade--off is a consequence of the need to balance the removal of small scale noise by smoothing the FSLE field and the need to avoid artificially reducing the FSLE gradients that are necessary to identify the strongest ridges. The results are not optimal yet and we are working on improvements to the extraction algorithm to reduce the impact of this trade-off.

Although the main focus of this analysis was the horizontal transport, vertical motions also play a role in the physical/biogeochemical interactions.  For instance, \citet{Omand2015} showed that the downward transport of surface carbon was
indeed favoured by the subduction of a filamental structure associated with mesoscale eddies. \deleted{This work opens new opportunities to detect
and fully characterize these structures and their impacts on tracers. Nevertheless, studying 3d transport of fluid and tracers in the ocean
still remains a challenge due to the predominance of horizontal stretching over vertical motions and the co-existence of diabatic and
adiabatic processes. Further work is needed to continue the development of Lagrangian indicators of 3d transport and mixing in geophysical
flows.}

\section{General conclusions}
\label{sec:c6-conclusions}

\added{In this work we investigated the full 3-dimensional structure as well as the cross-shore transport properties of an upwelling filament in the Iberian upwelling from a LCS perspective. Specifically, we showed that the filament can be characterized by its dispersion characteristics (its boundaries are formed by LCSs; low dispersion prevails in its interior). We also demonstrated that these dynamical properties are reflected in the hydrographic properties of the filament (LCSs match large gradients of temperature and salinity) allowing us to discuss the degree of isolation of water within the filament. This work opens new opportunities to detect and fully characterize 3d filamentary (sub)mesoscale oceanic structures and their impacts on the transport and dispersion of tracers. Nevertheless, studying 3d transport of fluid and tracers in the ocean still remains a challenge due to the predominance of horizontal stretching over vertical motions and the co-existence of diabatic and adiabatic processes. Further work is needed to continue the development of Lagrangian indicators of 3d transport and mixing in geophysical flows.}

\acknowledgments
The authors acknowledge support from Ministerio de Economia y Competitividad and Fondo Europeo de
Desarrollo Regional through the LAOP project (CTM2015-66407-P, MINECO/FEDER)
and through a Juan de la Cierva { Incorporaci\'on fellowship (IJCI-2014-22343) granted
	to V.R.
	J.H.B. acknowledges financial support from the Portuguese FCT
	(Foundation for Science and Technology) and Fundo Social Europeu
	(FSE/QREN/POPH) through the predoctoral grant SFRH/BD/63840/2009.
Data used in this study can be accessed at https://doi.org/10.5281/zenodo.802324.

\listofchanges

\end{document}